\documentclass[12pt]{article}

\usepackage{graphicx}
\usepackage{epstopdf}
\usepackage{amsmath}
\usepackage{amsfonts}
\usepackage{amssymb}
\usepackage{enumerate}
\usepackage{amsthm}
\usepackage{array}
\usepackage{epsfig}
\usepackage[small,bf]{caption}
\usepackage{appendix}
\usepackage{dsfont}
\usepackage{hyperref}
\def\1ad{\mbox{\normalsize $^1$}}
\def\2ad{\mbox{\normalsize $^2$}}
\def\3ad{\mbox{\normalsize $^3$}}
\def\4ad{\mbox{\normalsize $^4$}}
\def\5ad{\mbox{\normalsize $^5$}}
\def\6ad{\mbox{\normalsize $^6$}}
\def\7ad{\mbox{\normalsize $^7$}}
\def\8ad{\mbox{\normalsize $^8$}}

\def\beq{\begin{equation}}                     %
\def\eeq{\end{equation}}                       %
\def\bea{\begin{eqnarray}}                     
\def\eea{\end{eqnarray}}                       
\def\dj{\hbox{d\kern-0.347em \vrule width 0.3em height 1.252ex depth
-1.21ex \kern 0.051em}}

\def\half{{1\over 2}\,}

\def\ket{\rangle}
\def\bra{\langle}

\def\pt{\partial}

\def\shalf{{\mbox{$\half$}}}

\def\Dirac{\,\raise.15ex\hbox{/}\mkern-13.5mu D}
\def\dirac{\,\raise.15ex\hbox{/}\kern-.57em \partial}
\def\pslash{\,\raise.15ex\hbox{/}\kern-.57em p}

\def\CO{{\cal O}}
\def\CH{{\cal H}}

\begin{document}

                     %

\newcommand{\sheptitle} {Fast Scramblers, Horizons and Expander Graphs} \newcommand{\shepauthora} {{\sc Jos\'e
    L.F.~Barb\'on and Javier M. Mag\'an}}

\newcommand{\shepaddressa} {\sl
  Instituto de F\'{\i}sica Te\'orica  IFT UAM/CSIC \\
  Facultad de Ciencias C-XVI \\
  C.U. Cantoblanco, E-28049 Madrid, Spain\\
  {\tt jose.barbon@uam.es}, {\tt javier.martinez@uam.es} }

\newcommand{\shepabstract} {
  \noindent

 We propose that local quantum systems defined on expander graphs provide a simple microscopic model for thermalization on quantum horizons. Such systems are automatically fast scramblers and are motivated from the membrane paradigm by a conformal transformation to the so-called optical metric.}

\begin{titlepage}
  \begin{flushright}
    {IFT-UAM/CSIC-12-40\\
    }

\end{flushright}
\vspace{0.5in} \vspace{0.5in}
\begin{center} {\large{\bf \sheptitle}}
  \bigskip\bigskip \\ \shepauthora \\ \mbox{} \\ {\it \shepaddressa} \\
  \vspace{0.2in}

  {\bf Abstract} \bigskip \end{center} \setcounter{page}{0} \shepabstract
\vspace{2.7in}
\begin{flushleft}
  \today
\end{flushleft}


\end{titlepage}

\newpage


\setcounter{equation}{0}

\section{\label{intro} Introduction}

\noindent

Black hole horizons are conjectured to be maximally chaotic, in the sense of maximizing the rate of information scrambling \cite{Hayden, Sekino, Susskindnew}. Thermalization times for sufficiently localized near-horizon processes are conjectured to tightly approach causality bounds, i.e. the information-scrambling time over  a horizon patch of length $L$ would scale  logarithmically with the law
\beq\label{loglaw}
\tau_s (L) \sim \beta\,\log\, \left({L\over \ell_P} \right)\;,
\eeq
where $\beta$ is proportional to the  inverse Hawking temperature and time is measured in an appropriately red-shifted, off-horizon frame. The arguments supporting (\ref{loglaw}) are largely heuristic, based on no-cloning {\it gedankenexperiments} and ideas drawn from the `membrane paradigm' of stretched horizons \cite{Damour, Thorne, Susskindbook}. 

In \cite{usuno, usdos} we have examined these heuristic arguments to conclude that  they all depend on a simple kinematical criterion. Namely  $\tau_s$ in (\ref{loglaw}) is given by the free-fall time scale to the stretched horizon, across the Rindler region of the near-horizon geometry. This means that, once $\beta$ is fixed,   $L$ is always bounded, so that `fast scramblers' are necessarily small systems. As explained in \cite{usuno, usdos}, the  most general form of (\ref{loglaw}), applying to all kinds of horizons representing thermal states of hypothetical holographic dual systems, is given by
\beq\label{enteff}
\tau_s \sim \beta\,\log \,S_{\rm eff}
\;,
\eeq
where $S_{\rm eff}$ is the specific entropy, depending on the extensive  properties  of the system. For systems smaller than a single thermal cell (the length scale set by the inverse temperature $\beta= 1/T$), the specific entropy is jus the total entropy: $S_{\rm eff} =S$. An example of such behavior is a near-extremal Reissner--N\"ordstrom black hole. On the other hand,  for `large' systems, $S_{\rm eff}$ is defined as the entropy per thermal cell; $S_{\rm eff} = S_{\rm cell} = S/ VT^d$,  where $VT^d$ is the number of thermal cells in a volume $V$. A natural example would be a large black-brane horizon. Notice that the Schwarzschild black hole is a special borderline case, whose horizon size is roughly one thermal cell. 

The characterization of (\ref{loglaw}) as `fast' rests upon  the standard  intuition about information spread in local systems.  If information is codified in some microscopic degrees of freedom supporting a locally conserved charge, we  have transport phenomena associated to the conserved charge, and `scrambling' is interpreted as the diffusive component of such transport. At the classical level diffusion can be described by random walk behavior of quasiparticle degrees of freedom, leading to mixing time scales of order
\beq\label{difftime}
\tau_{\rm diff} (L) = D^{-1} \, L^2\;,
\eeq
where $D$ is the diffusion constant. For systems with $\CO(1)$ dimensionless parameters, we can estimate that $D\sim \beta$ is the  time scale defining the coarse-graining in space and time over which the random walk is defined.  At the quantum level, the interactions imply decoherence of any sufficiently localized subsystems where we may store the information. Hence, the classical diffusion picture should still apply provided the random-walk coarse graining scale is larger than the decoherence time scale. Thus, we expect (\ref{difftime}) to apply in order of magnitude over a wide range of conditions.  

Comparing (\ref{difftime}) and (\ref{loglaw}) at large $L$ we see the reason for the denomination `fast scrambler'. A better characterization  is obtained by
focusing on the entropic form (\ref{enteff}). If we evaluate (\ref{difftime}) for an extensive system with $(L/\beta)^d$ thermal cells, each one contributing an amount of  $\CO(1)$ entropy, we find 
\beq\label{diffen}
\tau_{\rm diff} \sim \beta\, (L/\beta)^2 \sim \beta \,S^{\,2/d}\;,
\eeq
which illustrates the `fast' character of (\ref{enteff}) in the limit of large entropy, as compared to an extensive `slow scrambler'  with the same number of degrees of freedom.

Our discussion in this paper assumes the usual tenets of `black hole complementarity' \cite{bhcompl}, namely the exterior description of the black hole can be regarded as quantum mechanically complete, notwithstanding the existence of other `complementary' observables which  would account for the `interior' physics. An effective model  for this `exterior description' is assumed where $A_{H}/4G$ degrees of freedom are approximately localized on the stretched horizon, a time-like surrogate of the event horizon with Planckian thickness.  If the law (\ref{loglaw}) is applied locally on the stretched horizon, one finds that it saturates causality \cite{susskindholo}, i.e. such a fast scrambling must operate at {\it ballistic} speed. This fact makes the conjectured fast-scrambling properties of horizons rather exotic and presumably require a  highly non-local effective dynamics for the stretched horizon, despite supporting naively the degrees of freedom of a local system with Planck cutoff (cf. \cite{fischler} for a recent example).

In this note we propose a middle ground in this conceptual dilemma. Borrowing some standard results in graph theory we show in section 2 that there exist local scramblers with exponentially fast mixing, satisfying (\ref{enteff}). The condition being that the random-walk's configuration space be an {\it expander graph} (cf. \cite{exprev}). We then conjecture that expander graphs are good local models of the stretched horizon, and we motivate this conjecture in section 3, using a reasoning analogous to ref. \cite{usuno}. In section 4 we briefly comment on the general entanglement properties of quantum systems defined on expander graphs, as well as more intrinsic characterizations of the scrambling. Finally we spell out our conclusions in section 5.

 \section{Local Scramblers {\it Can} Be Fast}
 
 \noindent
 
 A precise definition of information scrambling  in {\it finite} quantum systems touches upon many subtleties (information theoretic and dynamical alike, cf. \cite{haydennew} for a recent general discussion). For the purposes of this section, we start with a pragmatic and somewhat naive notion of `local scrambler'.  
 Namely  we regard as a scrambler any quantum system that can be considered as a `reservoir' for a reference `small' system which we call the probe.  Hence, the total Hilbert space reads 
 \beq\label{probhil}
  \CH = \CH_{\rm probe} \otimes \CH_{
\rm scrambler}
\eeq 
 and we say that the probed system is a  scrambler when the coupling between the two factors in (\ref{probhil}) randomizes the dynamics of the probe in the limit that ${\rm dim} (
 \CH_{\rm scrambler})$ is large,  i.e. we have an abstract  model of Caldeira--Legget type \cite{cl}. 
 
 We shall speak of local scramblers when a further dynamical condition is met. In general, the probe-scrambler coupling  selects a preferred decoherence basis on $\CH_{\rm probe}$, and we demand that the scrambler dynamics  be {\it local} in this basis. Namely, the scrambler can be modeled as  a local many-body quantum system over some lattice,  which is isomorphic to the decoherence basis of the probe. In this way, we model the probe dynamics as exercising `quantum brownian motion' over a lattice, driven by a many-body local system supported on the same lattice, so that ${\rm dim}(\CH_{\rm probe}) \sim \log \,{\rm dim} (\CH_{\rm scrambler})$. \footnote{It is important to keep in mind  that the `lattice' of decohered states is not necessarily tied to `real' space, and could arise as an effective description in, say the index space of a matrix model. }

 \subsection{Walks}
 \noindent
 
 Upon integrating out the scrambler degrees of freedom, the effective probe dynamics is described by a non-unitary master equation for the probe density matrix $\rho_{\rm probe}$. After an appropriate coarse-graining in space and time, the master equation often reduces to a discrete Markov process over the decoherence basis  of $\CH_{\rm probe}$ (cf. for example \cite{deco}). 

Let us represent a decohered probe state as a diagonal density matrix $\rho_{vw} = p_v \delta_{vw}$ in the decoherence basis, where the components $p_v$ satisfy positivity and normalization, $\sum_{v=1}^{N} p_v =1$, thus defining a probability distribution.  Pure basis states $|v_0 \ket \bra v_0 |$  are given by sharp distributions $p_v = \delta_{vv_0}$, whereas maximal mixing is represented by the uniform distribution $u$, defined by the components 
$$
u_v ={1\over N}\;.
$$
The Markov process is   determined by a  $N\times N$ matrix $M_{vw}$ of  transition probabilities between pairs of such basis states, so that on the $(n+1)$-th step 
\beq\label{markovc}
p^{(n+1)} = M\,p^{(n)} \;.
\eeq
Microscopic reversibility requires $M$ to be symmetric. We can codify the matrix of transition probabilities between basis states by means of an abstract graph consisting of $N$ vertices, one for each basis state $|v\ket$, and  edges $(vw)$ between pairs of distinct vertices for each
 non-vanishing off-diagonal matrix element  $M_{vw} \neq 0$, $v\neq w$.
 
 With no essential loss of generality, we may impose the discrete analog of homogeneity, so that every vertex has the same degree of connectivity with the rest of the graph, a fixed coordination number  $k<N$. Further imposing the analog of isotropy, we assume that all transition probabilities out of a given vertex are equal to one another.  Finally, the probability of no jump from a given site is denoted $r\in [0,1]$, and controls the diagonal elements $M_{vv}$. All in all we have
\beq\label{Mat}
M= {1-r \over k} \,A + r\;,
\eeq
where $A$ is the adjacency matrix of a simple, $k$-regular graph of $N$ vertices $G_{N,k}$.  The adjacency matrix of such a graph  is known to be  symmetric with spectrum in the interval $[-k,k]$. The uniform distribution is always an eigenvector of  both $A$ and $M$
\beq\label{unitd}
A \,u = k\,u\;,\qquad M\,u = u\;,
\eeq
which is non-degenerate if and only if the graph is connected. The lower bound for eigenvalues of the adjacency matrix, $-k$, is attained if and only if the graph is bipartite. 
It is useful to define the  Laplacian matrix of the graph
\beq\label{laplace}
\bigtriangleup \equiv 1-k^{-1} \,A\;,
\eeq
a positive-definite matrix with eigenvalues on the interval $[0,2]$, which gives a discrete approximation to the continuum Laplacian operator $-\nabla^2$. The uniform distribution is a zero mode of the Laplacian, $\bigtriangleup u = 0$, and the second largest eigenvalue of $A$ defines the gap $\Delta$, i.e. the  lowest {\it positive} eigenvalue of $\bigtriangleup$. 

In terms of the Laplacian matrix, the Markov process can be rewritten as a discrete diffusion equation
\beq\label{dfou}
p^{(n+1)} - p^{(n)} = -(1-r) \bigtriangleup \,p^{(n)}\;,
\eeq
whose explicit solution is given by the discrete version of the heat kernel,
\beq\label{hkerd}
p^{(n)} = M^n \,p^{(0)} =\left(1-(1-r) \bigtriangleup\right)^n \,p^{(0)}\;,
\eeq
in terms of the initial state distribution  $p^{(0)}$. 

Hence, all scrambling properties, at least as far as they are parametrized by the decoherent (effectively classical) walk  of the probe, are determined by the adjacency matrix $A$ or, equivalently by the graph Laplacian matrix $\bigtriangleup$.

For our purposes, the main property of interest is the efficiency of the approach to the uniform distribution as a function of the size of the graph. Let us look at the norm-difference between the $n$-th step distribution and the uniform one
$$
| p^{(n)} - u | = | M^n p^{(0)} - u | = |M^n (p^{(0)} -u) | \;.
$$
Since  $\sum_v p^{(0)}_v =1$, it turns out that $p^{(0)} -u$ is always orthogonal to $u$, and  we can bound 
$$
| p^{(n)} - u | \leq |{\bar m}|^n \,| p^{(0)} -u |  \leq |{\bar m}|^n\;,
$$
where $|{\bar m}|$ is the largest eigenvalue (in absolute value) of the matrix $M-1$.  At this point, it will be technically convenient to work with sizable  no-jump probabilities $r\geq 1/2$ which ensure positive-definiteness of $M$, so that $|{\bar m}| = {\bar m}$ is the largest eigenvalue of $M-1$, without regard to sign. In particular, we make the conventional  choice $r=1/2$, known in the graph theory literature as the `lazy random walk'. We shall see further down the discussion that these choices are not too essential from the physical point of view. 

For the lazy random walk, ${\bar m} = 1- \shalf \Delta$ and the bound becomes
\beq\label{bound}
| p^{(n)} - u | \leq \left(1-\shalf \Delta \right)^n\;.
\eeq
In other words, the rate of approach to the uniform distribution is controlled by the Laplacian gap $\Delta$. 

Our main interest is the behavior of $\Delta$ as the graph becomes very large.  
As a simple example, consider the  case of 
 a regular  $d$-dimensional hyper-cubic lattice $({\bf Z}_{l})^d$ of toroidal topology, with  degree $k=2d$
and vertex size $N= l^{\,d}$. The Laplacian spectrum is
$$
\Delta_{\vec \delta} = {2\over d} \sum_{i=1}^d \left(1-\cos(2\pi \delta_i / l)\right)\;.
$$
with ${\vec \delta} \in ({\bf Z}_l)^d$ and  gap  $\Delta = 4\pi^2 / l^2 = 4\pi^2 / N^{\,2/d}$.  Hence, the $p^{(n)}$ distribution becomes exponentially close to uniform for $n\gg n_{\rm mix} =N^{\,2/d}$. \footnote{If we demand that $|p^{(n)} -u| < 1/N^{\alpha}$ for some $\CO(1)$ positive number $\alpha$, we actually need a slight overkill with respect to $n_{\rm mix}$, i.e. we require $n>  C N^{2/d}\,\log(N)$ with $C$ another $\CO(1)$ constant.}  Since $N\sim \log {\rm dim} \CH_{\rm scrambler}$ controls the entropy of the scrambler, we recover the diffusion time scale
$$
\tau_{\rm diff} \sim \beta\,n_{\rm mix} = \beta \,N^{\;2/d} \sim \beta \,S^{\,2/d}\;,
$$
as expected. 

\subsection{Expanders}
\noindent

Much more interesting is the case of random walks on graphs having the property that the gap remains bounded
above zero, uniformly in the limit $N \rightarrow \infty$. Such graphs  are known as {\it expanders} and feature prominently in graph theory, on account of their numerous applications. 

A central result of expander graph theory is precisely the exponentially fast mixing of random walks. In particular for our lazy random walk we can bound (\ref{bound}) by $1/N^{\alpha}$ with $\alpha >0$, by setting $n> C \,\log(N)$
where $C$ is positive numerical constant proportional to $-\alpha /\log(1-\Delta_\infty /2)$ and $\Delta_\infty >0$ is the  $N \rightarrow \infty$ limit of the gap.

Hence, we reach the main lesson of this section: perfectly local random walks can behave as fast scramblers provided the graph is  an expander. 

Not-so-lazy random walks with $r<1/2$, hopping on $A$-matrices  with sufficiently negative eigenvalues, can have transition matrices $M$ with large negative eigenvalues which control the mixing properties. The extreme case of this kind is a bipartite graph, for which the long-time walk just oscillates between the two components, while well-randomized in each one of them. Therefore, the lack of precise mixing is related in this case to exact or approximate discrete symmetries of the graph. We can handle these cases by defining scrambling in terms of a single  randomized  component, which is equivalent to  increasing the time coarse-graining to even time steps. 

We have defined expander graphs in terms of the spectral properties of the Laplacian. There is a geometrical
interpretation of the Laplacian gap, as a consequence of the   so-called Cheeger  inequality
\beq\label{cheeger}
{h^2 \over 2} \leq \Delta \leq 2h\;,
\eeq
where the Cheeger constant, $h$, is the (normalized) edge expansion of the graph, defined as the formal ratio of `area/volume', 
\beq\label{edgeexp}
h\equiv {\rm min} \; {{\rm \# \,edges} \,[\,\pt\,S\,] \over k\,N_S}\;,
\eeq
where the minimum is taken over all graph bi-partitions $G=S \cup (G-S)$, with the proper subgraph $S$ having vertex size $N_S$  smaller than half of $G$, i.e.  $N_S \leq N/2$.  In the numerator we count the total number of edges connecting the two sides of the partition, thus measuring the boundary volume in edge units. Therefore, an expander graph family has a bounded finite $h$ in the large $N$ limit, implying that large subsets have
a boundary volume scaling as their bulk volume. We recognize expanders as the discrete analog of hyperbolic geometry.

\begin{figure}
  \label{Bethe}
  \begin{center}
   \epsfig{file=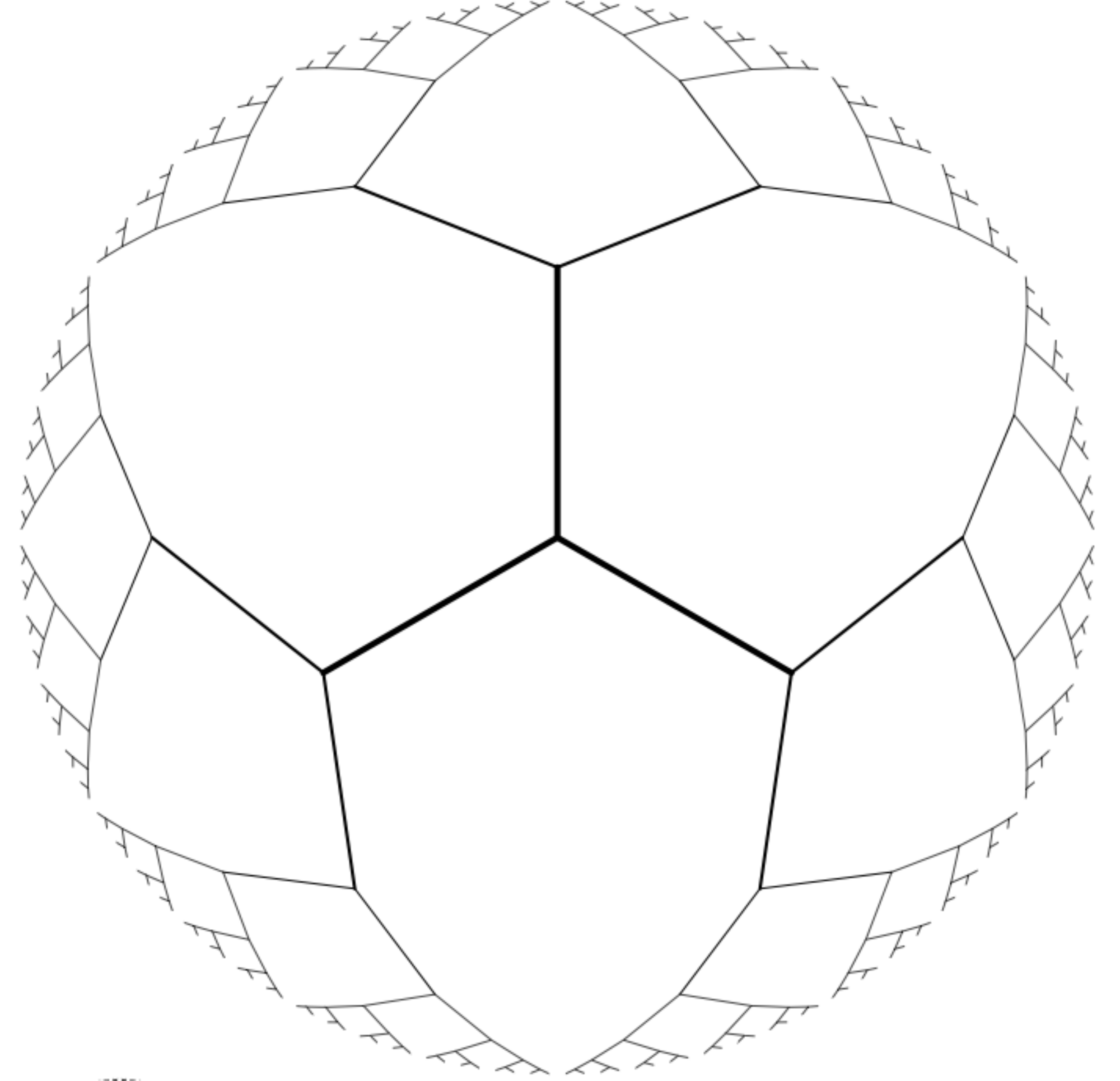, width= 5.5cm}
    \caption{ \small The infinite regular tree, or Bethe lattice, is the ultimate expander. Its adjacency matrix has a continuous spectrum in the interval $[-2\sqrt{k-1} , 2\sqrt{k-1}] $. The figure shows the case $k=3$, a tessellation of the two-dimensional hyperboloid. }
  \end{center}
\end{figure}

For example, the simplest and in many ways the most perfect expander is a regular tessellation \cite{tessellations}  of the Hyperbolic space  ${\bf H}^{k-1}$ known as the Cayley tree or Bethe lattice (cf. figure 1). The metrical properties  of the $k$-regular Bethe lattice give an approximation of the $(k-1)$-dimensional unit hyperboloid  
\beq\label{hypome}
ds^2_{{\bf H}^{k-1}} = dr^2 + (\sinh \,r)^2 \,d\Omega_{k-2}^2 \;,
\eeq
with lattice spacing of  order unity.\footnote{Continuum models of finite expander graphs can be constructed as compact hyperbolic manifolds of the form ${\bf H}^{k-1} /\Gamma$, with $\Gamma$ a free discrete subgroup of isometries.} We can introduce  hyperbolic spaces of arbitrary curvature radius
$R$, just scaling this metric by a factor of $R^2$, and the corresponding Bethe lattice would have spacing of $\CO(R)$.

The discrete diffusion equation 
(\ref{dfou}) is also reminiscent of the continuum diffusion equation (cf. \cite{brownhypo}):
\beq\label{contdiff}
\pt_t \,p = D\; \nabla^2 p\;,
\eeq
where $D$ is the diffusion coefficient, $p(t, x)$ is the continuum diffusing density  and $-\nabla^2  $ is the Laplacian. For the hyperbolic space with curvature radius $R$ it reads 
$$
 -\nabla_{\bf H}^2 =  -{1\over \left(\sinh (z/R)\right)^{k-2}}\; {d\over dz} \;\left( \sinh ( z/R)\right)^{k-2} \;{d \over dz} - \left({1\over R\sinh (z/R)}\right)^2\, \nabla^2_{{\bf S}^{k-2}}\;, 
$$
where $-\nabla^2_{{\bf S}^{k-2}}$ is the Laplacian on the unit sphere, $R$ is the curvature radius of the hyperbolic space and we have defined $z= r R$. 

To investigate the speed of diffusion we look at  the heat kernel
 $$
 K(z, \Omega, t)= \bra z, \Omega | \,\exp \left(t\,D^{-1}\,\nabla^2_{\bf H}\right) | 0 \ket
 $$
  describing the diffusion of a distribution which begins  sharply localized at the origin $z=0$.  Since that distribution is spherical, we may rewrite the diffusion equation as a one-dimensional equation for the radial density 
 $$
 f(z,t )= K(z, t)\left( \sinh (z/ R)\right)^{k-2}\;,
 $$
  which reads
$$
\pt_t \,f +  \pt_z\, \left( v \cdot f\right) = D\, \pt_z^2\, f \;.
$$
This equation  has the Fokker--Planck form with a drift velocity 
$$
v(z )={(k-2)\over \tanh (z/R)} {D \over R}\;,
$$
which approaches exponentially fast the limiting value $v_\infty = (k-2) D /R$ as $z\gg R$.

In comparing to the discrete diffusion over an expander graph, we recall that the lattice spacing is or the order of the curvature radius, and assuming the free-flight velocity is of $\CO(1)$ both are of the order of the time step $\beta$. Hence the continuum problem approximates the discrete one for $D\sim R = \CO(\beta)$. 
 The drifted 
radial distribution is peaked around $z_{\rm peak} (t) \sim v_\infty  \,t + \CO(\sqrt{t\beta}\,)$, with $v_\infty $ of order one, implying a radial spread of  {\it ballistic} type for the hyperbolic diffusion.

 The same ballistic character can be argued directly in the discrete random walk on an expander graph. Defining discrete spheres as the set of vertices at edge minimal distance $n$ from a fixed central point, and balls $B_n$ as the interior of those spheres, we can reason as follows. Equation (\ref{edgeexp}) implies that the number of edges coming out of a ball $B_n$ is larger than $h k N_n$
 where $N_n$ is the vertex size of $B_n$. Since at most $k$ edges can end on a vertex, the number of vertices
 in the $B_{n+1}$ ball is at least $N_n + h N_n = (1+h) N_n$. Thus $N_n \geq (1+h)^n$ and we see that  the vertex volume of concentric balls grows exponentially. This means that the size of a $N$-vertex ball is of order $\log \,N$, the same as the random-walk time for mixing on this sphere, showing that the spread of the walk is indeed ballistic.

\section{The expander at the edge of the horizon}

\noindent

Having discussed how abstractly defined local scramblers can be {\it fast}, we now come back to original black hole problem. An intriguing connection arises between expanders and black-hole horizons  through the universal properties of the near-horizon geometry and in particular the natural description of stretched horizons in the so-called {\it optical frame}, obtained by removing the $|g_{00}|$ factor in any static metric by a conformal transformation (cf. for example \cite{opticalnew} for a recent review). 

\subsection{Expanders and de Sitter}
\noindent

The basic observation starts from the geometrical properties of expander graphs described in the previous section, namely the effective description of brownian motion as hyperbolic diffusion. An immediate connection to a gravitational system is found by noting the global conformal map between the space ${\bf R} \times {\bf H}^n$ and the static patch of de Sitter space-time, ${\rm dS}_{n+1}$, with metric
$$
ds^2_{\rm dS} = (1-u^2) \,dt^2 + (1-u^2)^{-1} du^2 + u^2 \,d\Omega_{n-1}^2\;.
$$
The corresponding optical frame is given by
$$
ds^2_{\widetilde {\rm dS}} = -dt^2 + {du^2 \over (1-u^2)^2} + {u^2 \over (1-u^2)} \,d\Omega_{n-1}^2 = -dt^2 + dr^2 + (\sinh r)^2 \,d\Omega_{n-1}^2\;,
$$
which is globally identical to ${\bf R} \times {\bf H}^{n}$ under the coordinate change $\sinh r = u(1-u^2)^{-1/2}$. 
Under the conformal map, null trajectories on dS$_{n+1}$ are mapped to null trajectories on ${\bf R} \times {\bf H}^n$, which project to geodesic arcs on ${\bf H}^n$. Therefore, if the hyperbolic diffusion is realized as a random walk with piecewise geodesic arcs of unit length (in units of the curvature radius), it maps to a random walk on the static patch of de Sitter space-time made out of piecewise-null free paths. In other words, if we model scrambling on the static dS patch as the piecewise-null scattering of a localized probe with free path time scale given by the curvature, such a system behaves as a fast scrambler.

If the hyperbolic random walk is regularized to $N$ vertices, the whole system is scrambled in a time of order $\log\,N$. Since the vertex density is of $\CO(1)$ in units of the hyperbolic curvature, $N$ also scales as the hyperbolic volume, which in turn scales like the boundary `area'. Thus, restoring units we have a time scale of order  $H^{-1} \,\log S_{
\rm GH}$ with $H$ the Hubble parameter and $S_{\rm GH}$ the Gibbons--Hawking entropy of de Sitter space-time.

The understanding of hyperbolic geometry as the key to fast scrambling is also manifest in a different mechanical model for fast scrambling of dS space-time.  Let us suppose that the probe, instead of executing a brownian motion with time scale of $\CO(1)$ in units of the curvature,
 glides freely through dS space-time except for localized scattering at the Planckian stretched horizon, which is regarded as a set of $S_{\rm GH}$ scatterers with Planckian cross section. In order to derive a bound on scrambling speed, we take the probe glides between horizon scatterings as null paths on dS, so that the complete trajectory maps to a piece-wise geodesic trajectory on the optical frame. This defines a Hadamard billiard obtained by carving $S_{\rm GH}$ geodesic spheres out of ${\bf H}^n$, with cross section of $\CO(1)$ in units of the curvature radius \cite{billiard}. The scrambling time for a probe with collision resolution of the order of one scatterer  in such a billiard is given by the crossing time of the billiard, i.e. its diameter $\tau_s \sim \beta\, \log\, S_{\rm GH}$, where we have restored the radius of curvature $\beta$.

We thus make contact with the chaotic scrambling models of \cite{usuno}, exhibiting the Hadamard billiard probe-scattering  and the expander graph probe-diffusion as two extreme models of fast scrambling in the de Sitter static patch, depending on whether we confine probe interactions strictly to the Planckian stretched horizon, or we conceive of an `extended' stretched horizon covering the whole dS patch, as far as probe interactions concerns.

\subsection{Local analysis}
\noindent

The previous discussion exhibiting the global map between the de Sitter static patch and the hyperbolic geometry can be generalized to a local approximate map for any regular horizon. The basic observation is again the conformal optical map between ${\bf R}\times {\bf H}^n$ and
Rindler space-time, with metric
$$
ds^2_{\rm Rin} = -y^2 dt^2 + dy^2 + d\ell^2\;,
$$
where the $\ell$ coordinates parametrize ${\bf R}^{n-1}$. The optical frame then yields ${\bf R} \times {\bf H}^n$ again, presented now in  the upper-half plane coordinates
$$
ds^2_{\widetilde {\rm Rin}} = -dt^2 + {dy^2 + d\ell^2 \over y^2}\;.
$$
Therefore, cutoff versions of Hadamard billiards or expander graphs on the hyperbolic upper-half plane representation yield fast-scrambling models for Rindler space-time, which in turns gives a local approximate description of any regular horizon.

To see this, notice that near a regular, static horizon we may choose $(d+2)$-dimensional coordinates exhibiting a two-dimensional Rindler factor in the near-horizon geometry ${\bf X}^{d+2}$,  of the form
\beq\label{nearhor}
ds_{\bf X}^2 =g_{\mu\nu} \,dx^\mu dx^\nu \approx F(\rho, y)\left(-\left({2\pi \rho\over \beta}\right)^2\, dt^2 + d\rho^2\right) + H_{ij} (\rho, y) \, dy^i \,dy^j\;,
\eeq
where the horizon is conventionally located at $\rho=0$. The functions $F(\rho, y)$ and $H_{ij} (\rho, y) $ are assumed to have  smooth expansions around $\rho=0$. The normalization of the time-time component takes into account the fact that the Hawking temperature is assumed to be $1/\beta$, so that the imaginary-time periodicity $t\equiv t+ i\beta$ is equivalent to the absence of conical singularity in the Euclidean section.

The optical manifold ${\widetilde{\bf X}}$ is defined in the exterior region $\rho>0$ through the conformal transformation 
\beq\label{op}
ds^2_{\;\bf X} = \Omega^2 (\rho, y) \,ds^2_{\;{\widetilde{\bf X}}}  \;, \qquad \Omega (\rho, y) = {2\pi \rho \over \beta} \sqrt{F(\rho, y)}\;,
\eeq 
with near-horizon asymptotic  form 
\beq\label{opticalm}
ds^{2}_{\;\widetilde{\bf X}} \approx -dt^2 + dz^2 + e^{4\pi z / \beta}\; \gamma_{ij} (y)  \,dy^i dy^j \;, 
\eeq
where $\gamma_{ij} = (H_{ij} / F)_{\rho=0} + \CO(\rho^2)$,  and we have defined a Regge--Wheeler or `tortoise'  radial coordinate $\rho = e^{-2\pi z /\beta}$, in terms of which the Rindler region extends roughly along the interval $z\in [0,\infty]$, with the horizon sitting at $z=\infty$. 

Up to terms exponentially suppressed at large $z$, the near-horizon optical metric is `asymptotically locally hyperbolic', with obvious `$z$-expanding' properties in the sense of the last section. 
The spatial metric $d\ell^2 = \gamma_{ij} \,dy^i dy^j$ is induced at the edge of the Rindler region, and it is related to the physical induced metric at the horizon by a regular conformal transformation. Since $\gamma_{ij}$ defines a conformal compactification of the horizon boundary, the obvious analogy with the AdS/CFT correspondence was exploited in Ref. \cite{solo} to propose a dual Euclidean CFT description on the $\gamma$-geometry. In this paper, we take a different route and emphasize the  relation between (\ref{opticalm}) and microscopic models of the stretched horizon.

\subsection{The optical stretch}
\noindent

In this section we explain why the optical metric is a useful tool in any discussion of near-horizon dynamics. In general, in the context of `black hole complementarity' \cite{bhcompl}, we shall define
the stretched horizon as the limit of applicability of Low-Energy-Field-Theory (LEFT). Any such description involves a finite number of local fields  with Hartle--Hawking boundary conditions at $\rho=0$, which are equivalent to the specification of a thermal state at locally measured  temperature $(\beta \sqrt{-g_{00}})^{-1}$. Let us parametrize the effective action of any such LEFT in the schematic form (here for a scalar field degree of freedom $\Psi$)
\beq\label{lefta}
S_{\rm LEFT} =- \int_{{\bf X}^{d+2}} \,\left[\shalf \Psi \left( -\nabla^2_{\,\bf X} + {d \over 4(d+1)} R[\,{\bf X}\,]+  M^2 \right) \Psi + \sum_{\rm irr\; \CO} {\lambda \over \Lambda^{\Delta-d-2}} \CO(\Psi) \right]\;.
\eeq
In this expression, the Laplacian  has been corrected by a conformal coupling to the Ricci-scalar $R[{\bf X}]$, so that the remaining terms give the strength of conformal symmetry breaking. $M$ is a typical mass, standing  as a representative of whatever relevant  operators we may have turned on, and the remaining sum accounts for characteristic  irrelevant operators with conformal weight $\Delta > d+2$, dimensionless coupling $\lambda$ and scale $\Lambda$, playing the role of  Wilsonian UV cutoff for the LEFT.\footnote{We adopt the `naturality' convention in which the strength of the coupling is controlled by $\Lambda$, i.e. $\lambda = \CO(1)$.} Consistency of the Wilsonian procedure requires $M< \Lambda$. Since
the Hartle--Hawking quantum state to be stored on the fields $\Psi$ has energy scale $1/\beta$, the Wilsonian consistency also requires the constraint $1/\beta < \Lambda$.

At this point we are free to perform field redefinitions in (\ref{lefta}) if they serve the purpose of illustrating the physics, just as it is sometimes natural to switch back and forth between so-called Einstein and string-frames in the graviton-dilaton system arising in low-energy string-theory  Lagrangians. In the same vein, we shall rewrite (\ref{lefta}) in the `optical' frame defined by the conformal transformation (\ref{op}), together with the field redefinition 
$$
\Psi = \Omega^{-d/2}\,{\widetilde \Psi}\;.
$$
Using the identity 
$$
\left(-\nabla^2_{\,{\bf X}} + {d\over 4(d+1)} R[\,{\bf X}\,]\right) \Omega^{-d/2} = \Omega^{-2-d/2} \,\left(-\nabla^2_{\,{\widetilde {\bf X}}} + {d\over 4(d+1)} R[\,{\widetilde {\bf X}}\,] \right)
\;,
$$
we finally obtain
\beq\label{oplefta}
S_{\rm LEFT} =- \int_{{\widetilde {\bf X}}^{d+2}}\,\left[ \shalf {\widetilde \Psi} \left(- \nabla^2_{\,{\widetilde {\bf X}}} + {d \over 4(d+1)} R[\,{\widetilde {\bf X}}\,] + M^2_{\rm eff} \right)\,{\widetilde \Psi} + \sum_{\rm irr\; \CO} {\lambda \over (\Lambda_{\rm eff})^{\Delta-d-2}} \CO( {\widetilde \Psi}) \right]\;.
\eeq
The result is a conformally-invariant effective action, perturbed by relevant and irrelevant operators with effective position-dependent dimension-full parameters  $M_{\rm eff} (z)= \Omega(z) M$ and $\Lambda_{\rm eff} (z) = \Omega(z) \Lambda$. Since $\Omega(z\rightarrow \infty)\rightarrow 0$, all these dimension-full parameters are scaled to zero in the horizon limit. Hence, the $z\rightarrow\infty$ limit is a zoom into
the UV structure of the LEFT, isolating the  conformally-invariant dynamics of whatever field degrees of freedom exist below the Wilsonian cutoff $\Lambda$, while at the same time blowing up the irrelevant operators heralding the approach to the UV cutoff.\footnote{The conformal mass term is {\it negative}, approaching $-(\pi\, d/\beta)^2$ as $z\rightarrow\infty$. Since the ${\widetilde {\bf X}}$-Laplacian is gapped in the same amount, only the zero mode of ${\widetilde \Psi}$ is tachyonic. This mode is seriously non-normalizable in the original ${\bf X}$-frame field $\Psi$, so it does not satisfy the Hartle--Hawking boundary conditions and  should be treated as non-dynamical in the LEFT.} Since the thermal state of interest has scale $1/\beta$, the running of the effective dimensionless coupling   of  irrelevant operators, defined at scale $1/\beta$, is given by 
\beq\label{strongc}
\lambda_{\rm eff} (z) = {\lambda \over (\beta \Lambda \Omega(z))^{\Delta-d-2}}\;.
\eeq
Since $\lambda = \CO(1)$, the effective position-dependent coupling  enters strong-coupling at the point $z_\Lambda$ determined by 
\beq\label{lambdastretch}
\Omega(z_\Lambda) \Lambda = 1/\beta\;.
\eeq
Therefore, $z=z_\Lambda$ marks the end of the LEFT domain of applicability, and thus defines the edge of the  {\it $\Lambda$-stretched} horizon. The minimal choice for $\Lambda$ is on the order of the Planck mass, $M_{\rm P}$, 
defining the standard notion of a Planckian stretched horizon at the corresponding solution $z_{\rm P}$ of (\ref{lambdastretch}), but other choices are possible. Consider for example a perturbative string theory 
 with string mass scale $M_s$, which defines a stringy stretched horizon at optical depth $z_s$. In general we have $\Lambda = M_s = g_s^{\;\alpha} M_{\rm P}$ with $g_s \leq 1$ the string coupling and $\alpha>0$.  Solving the equations $1/\beta = \Omega(z_s) M_s = \Omega(z_{\rm P}) M_{\rm P}$ yields the optical depth difference
 $$
 z_{\rm P} - z_s = {\beta \over 2\pi} \log \left({M_{\rm P} \over M_s}\right) = {\beta \over 2\pi} \,\log(1/g_s^\alpha)\;.
 $$
 In this case, despite the fact that the string theory remains weakly coupled at $z\sim z_s$, the effective mass of the whole Hagedorn tower of massive string states comes down below the $1/\beta$ scale as $z<z_s$. Therefore, the dynamics is expected to become non-perturbative because of the exponential growth of field species.  
 
 Another interesting case is that of a `thick' stretched horizon, arising for $\Lambda \sim 1/\beta$, i.e.  a situation where there is essentially no LEFT description at all. This corresponds to the bulk dynamics of holographic gauge theory duals with 't Hooft coupling of $\CO(1)$. Such models are akin to large-$N$, weakly coupled matrix models with possibly complicated interactions.

The physics of (\ref{lefta}) in the radial domain $0\ll z\ll z_\Lambda$,  deep within the Rindler region but still clear from the stretched horizon, consists of a conformal thermal state. \footnote{The discussion leading to (\ref{lefta}) generalizes to any non-trivial CFT perturbed by relevant and irrelevant operators with respective scales $M$ and $\Lambda$.} Since the optical metric is ultra-static, the thermodynamic functions of such a thermal state are extensive on the hyperbolic spatial geometry. For example, the  entropy of $\Psi$-field states contained in Rindler, up to optical depth $z$,  scales with the optical volume 
\beq\label{entoex}
S_{\Psi} \sim {\widetilde V}_z \,\beta^{\,-d-1}
\;,
\eeq
where 
\beq\label{opvol}
{\widetilde V}_z = V \,\int_0^z dz' \,e^{2\pi z' d/\beta} \approx V\,{e^{2\pi z d/\beta} \over 2\pi d}\;,
\eeq
and $V\equiv V_\gamma$ is the volume in the $\gamma_{ij}$ metric. Hence, we have a uniform excitation density of $\CO(1)$ degrees of freedom per thermal volume  $\beta^{\,d+1}$ of optical bulk, although the `expanding' nature of the geometry forces  all extensive quantities to become dominated by the large $z$-cutoff (states pile up in the last layer of $\CO(\beta)$ thickness). The old result of 't Hooft's  brick-wall model \cite{brickwall} follows in this language as the (approximate) matching  of the entropy in  sub-Planck Hawking radiation with the Bekenstein--Hawking entropy of the black hole: 
\beq\label{bri}
S_{\rm BH} \sim {\widetilde V}_{z_{\rm P}} \,\beta^{\,-d-1}\;.
\eeq
Again, this entropy is roughly accounted for by the last layer of  optical thickness of  $\CO(\beta)$ (equivalently $\CO(\ell_{\rm P})$ thickness in the physical metric). 

\subsection{The optical expander}

\noindent

We will now make a basic assumption regarding the structure of the $\Lambda$-stretched horizon. We declare that the geometrical structure of ${\widetilde{\bf X}}$ is still relevant to characterize the interactions of strongly-coupled degrees of freedom building up the stretched horizon. 
More precisely, we model  the stretched horizon  as an effective hyperbolic lattice of $\CO(\beta)$ spacing with a worth of $\CO(1)$ Q-bit degrees of freedom per lattice point. In other words, we have a strongly-coupled  discrete quantum system on a cut-off expander graph.

 The dimension of this Hilbert space is fixed by the matching to the Bekenstein--Hawking entropy. In particular, we have the extensive behavior (\ref{entoex}), now applying to  the dimension of the Hilbert space,  rather than simply the entropy of excitations:
\beq\label{mat}
\log {\rm dim}\,\CH_{\rm SH} \sim {\widetilde V}_{\rm SH} \,\beta^{-d-1} \sim S_{\rm BH}\;,
\eeq
where ${\widetilde V}_{\rm SH} \sim {\widetilde V}_{z_{\rm P}}$ and we have used  (\ref{bri}).

Setting  $V=V_\gamma = \beta^{\,d}$, we deal with a single thermal cell, and the associated entropy is just the specific entropy $S_{\rm eff} = S_{\rm cell} =N$. Inverting  (\ref{opvol}) we find the relation between the maximal depth of the stretched horizon and the specific entropy
\beq\label{matchzetap}
z_{\rm P} = {\beta \over 2\pi d} \,\log \, N\;,
\eeq
for the total optical depth at the `bottom'  of the stretched horizon.

At this point, we can make contact with the discussion in section 2, characterizing horizon scrambling in terms of a discrete diffusion process with step $\beta$, playing on the expander graph cutoff to depths $z_\Lambda < z<z_{\rm P}$. 
The expander-diffusion process scrambles $\CO(n)$ degrees of freedom of a single thermal cell in a time of order $\beta \log(n)$, as the random walk reaches the `bottom' of the stretched horizon at $z=z_{\rm P}$. Any subsequent scrambling over scales larger than a thermal cell of the $\gamma_{ij}$ metric proceeds by standard diffusion on the `bottom' of the expander with induced (optical) metric
\beq\label{botme}
d\ell_{z_{\rm P}}^2 = e^{4\pi \,z_{\rm P} /\beta} \,\gamma_{ij} \,dy^i dy^j = N^{2/d} \,\gamma_{ij} \,dy^i dy^j
\;.
\eeq
Hence, a length scale $L\gg \beta$ in the $\gamma_{ij}$ metric corresponds to a length scale $L \,N^{1/d} \sim (NV)^{1/d}$ in the bottom metric (\ref{botme}). Such a patch   is slow-scrambled in a time of order $\beta^{-1} \,(V N)^{2/d}$. The number of thermal cells on the patch is $V/\beta^d$. Since $N$ is the entropy per thermal cell we may write the total entropy of the patch as $S \sim N \,V\,\beta^{-d}$, so that the total slow-scrambling time is the standard diffusive 
 $\beta \,S^{\,2/d}$ result, as in (\ref{diffen}). It is important to notice that this result is non-trivial, since it has been derived for systems which have $\CO(N)$ degrees of freedom per thermal cell, rather than $\CO(1)$ degrees of freedom as was the case in the Introduction.

Combining these estimates we finally find the fast-scrambling law 
\beq\label{scramfin}
\tau_s (L) \sim \beta\,\log\,N \;, \qquad L\leq \beta
\eeq
below the thermal cell scale, and the slow-scrambling law
\beq\label{slowsf}
\tau_s (L) \sim \beta \,S^{\,2/d}\;, \qquad L \gg \beta
\eeq
for distances well above the thermal cell.

\begin{figure}
  \label{Bethe}
  \begin{center}
   \epsfig{file=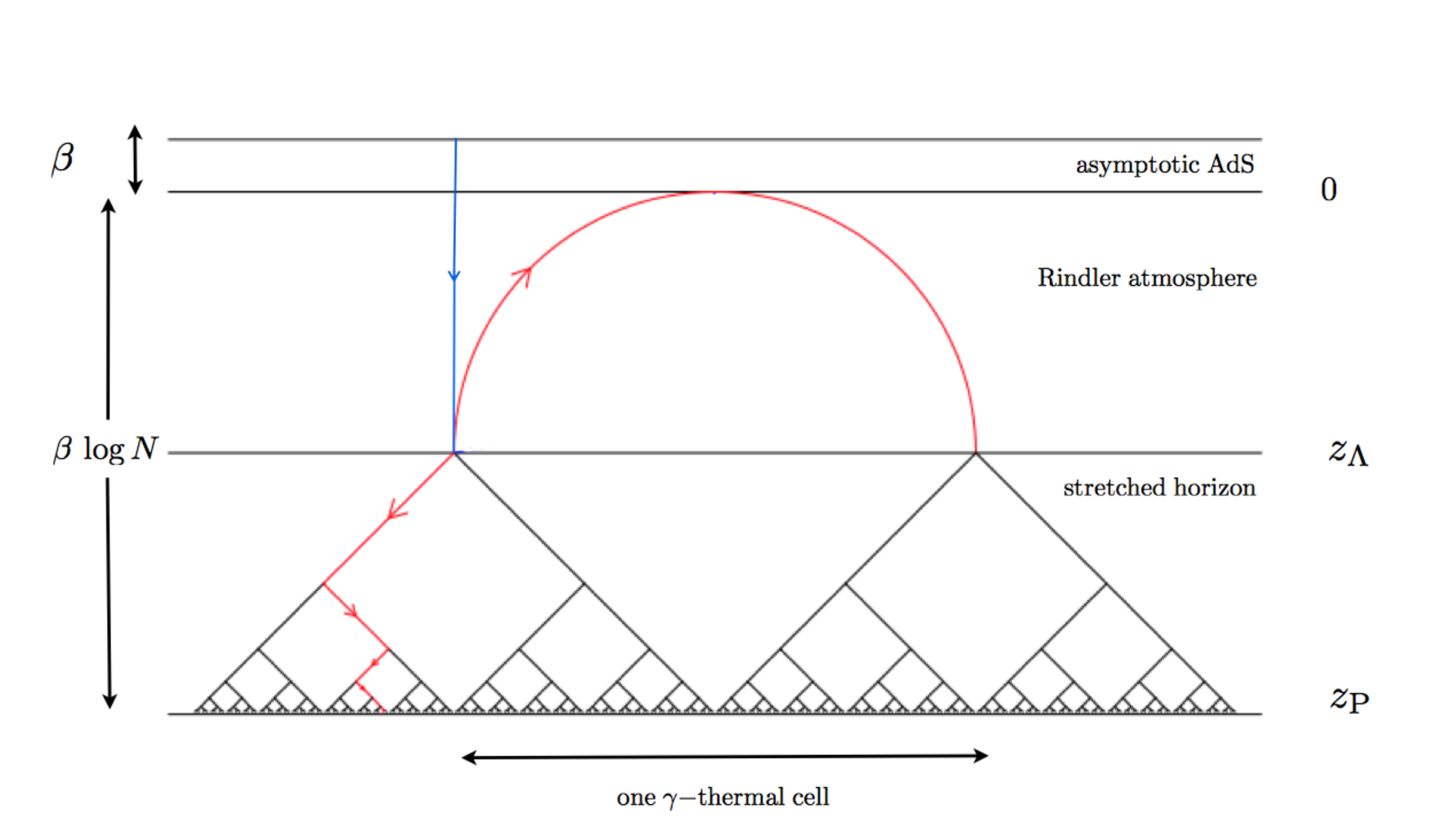, width= 12.5cm}
    \caption{ \small Schematic view of the effective kinetic scrambling model in the optical frame, featuring a Cayley tree picture of the stretched horizon, a hyperbolic `Rindler atmosphere'
    and an asymptotic AdS region of optical depth of $\CO(\beta)$. The total optical depth of the hyperbolic section $[0, z_{\rm P}]$  is $\CO(\beta\log N)$. A localized classical probe injected from the asymptotic boundary ( vertical blue line) is either absorbed  or reflected at $z=z_\Lambda$. If reflected (curved red line), it scrambles fast in the effective chaotic billiard of the Rindler atmosphere. If absorbed (jagged red line), it scrambles fast by diffusion on the expander graph until it reaches the bottom at $z_{\rm P}$. Once at the bottom, it scrambles slowly by diffusion on the bottom. The fast-scrambling patch measured by the
    $\gamma_{ij}$ metric is always about one thermal cell. }
  \end{center}
\end{figure}

The estimate (\ref{slowsf}) is quite general, since it only depends on the bottom density of the expander. On the other hand, (\ref{scramfin}) assumes tacitly that the expander graph starts at $z= z_\Lambda \sim 0$, i.e. the stretched horizon covers the whole Rindler region. Such a situation is only appropriate for highly stringy  black holes dual to weakly-coupled CFTs.  Smooth horizons with a large Rindler region where LEFT is valid support `thin' stretched horizons for which the fast-scrambling diffusion operates for a smaller time of order $\beta \log (N) - z_\Lambda$. In particular, in the limit in which we push the LEFT validity down to Planckian distances from the horizon, we have $z_\Lambda \sim z_{\rm P}$ up to $\CO(1)$ factors, and the optical thickness of the stretched horizon is only  of $\CO(\beta)$. In such a situation the random walk diffusion is always slow-scrambling. However, as mentioned above and argued in \cite{usuno}, surface-scattering effects turn the LEFT region $0<z<z_{\rm P}$ into a chaotic billiard with Lyapunov time of order $\beta\,\log N$. Therefore, the combined effect of classical chaos with
$\CO(1)$ `atmospheric' collision events and the standard diffusion upon absorption by the  expander graph account for the expected scrambling properties of idealized probes. A cartoon of this picture is offered in figure 2.

\section{Beyond the probe approximation}

\noindent

Up to this point  have modeled scrambling in a rather naive fashion, using the random walk model. Our main contention is the ability to perform ballistic-fast scrambling in this way, provided we define an appropriate background geometry for the diffusion process. This picture requires the operational definition of a probe system which is stable along the time evolution of the system and effectively classical on a coarse-graining time scale $\beta$. It would be interesting to free the discussion from this requirement, in such a way that the fast-scrambling property is characterized in a more fundamental way. 

In this section we advance some preliminary remarks about thermalization in quantum systems defined on expander graphs. We make no  explicit reference to a probe subsystem or any other quasiparticle degree of freedom. Rather, we focus on generic properties of entanglement  with respect to bi-partitions of the expander graph, and point out some difficulties in arriving at a good physical definition of scrambling. 

Let $G$ be a connected graph where a local quantum system is defined and $A \cup {\bar A}$  a bi-partition into two connected components, with the convention that $A$ has the smaller    vertex size:  $N_v (A) \leq N/2$. Given a general state on $G$, described by a density matrix $\rho$, we define the marginal density matrix of $A$ and its entanglement entropy as 
$$
\rho_A = {\rm Tr}_{{\bar A}} \,\rho\;, \qquad S_A = -{\rm Tr}_A\, \rho_A \,\log\,\rho_A\;.
$$
A natural criterion to declare a system as scrambled is to demand that $S_A $ be bounded below by some fixed {\it extensive} function, for all choices of $A$ under the stated conditions,
\beq\label{pagec}
S_A \geq C \,N_v (A)\;, \qquad N_v (A) \leq N/2\;.
\eeq
For systems with finite-dimensional Hilbert space at each vertex, this criterion is equivalent to maximal mixing of $\rho_A$ when the constant $C$ saturates to its upper bound  $C_{\rm max} = \log\dim \CH_v$. In other cases, it incorporates the fact that we may have different effective temperatures depending on the total energy of the state. Alternatively, one can speak of maximal mixing on the fixed energy surface, assuming ergodicity, i.e. no other conserved quantities besides the total energy.  The intuition behind the criterion is that proper subsystems see the complement as a thermal reservoir. The condition (\ref{pagec}) is expected to be generic \cite{page} for  randomly chosen global states $\rho$, and we shall refer to it as the Page criterion. 

The scrambling time may be defined as the typical lapse of Hamiltonian evolution needed  to satisfy the Page criterion, starting from an initial state which explicitly violates it. Of course, most of the difficulties in establishing general results lie on the lack of precision in just every concept appearing in the previous sentence. 

In order to sharpen the language a bit we may focus  on globally pure states on $G$. Then, a good choice for the `least thermalized' initial state may be a completely factorized one $\otimes_v |\psi_v\ket_v$, for which $S_A =0$ for all partitions. An up-to-date discussion of scrambling times for such initial states can be found in \cite{haydennew}. 

Another natural choice of initial state is one whose  entanglement properties imitate those of the ground state. For systems with short-range correlations, such
ground states satisfy the so-called `area law' for the entanglement entropy, i.e. $S_A \sim {\rm size} [\,\pt A\,]$, where ${\rm size} [\,\pt A\,] $ can be measured in terms of the edge size of the boundary, $N_\ell (\,\pt A\,)$, for $k$-regular graphs. The usual intuition about
local systems in Euclidean geometry suggests then that the scrambling consists on the gradual evolution from `area law' to `volume law' for the entanglement entropy. 

Already at the level of  these primitive notions, we see that expander graphs are special in their behavior with respect to {\it extensivity}. In particular, there is no clear-cut distinction between  `area-law' and `volume-law' behaviors when $G$ is an expander.       
As a consequence of the expansion property (\ref{edgeexp}), any area-law entanglement entropy of a bi-partition $G= A \cup {\bar A}$ satisfies the Page criterion
$$
S_A = s_A \, N_\ell (\,\pt A\,) \geq h\,k\,s_A \,N_v (\,A\,)\;
$$
from the beginning. Such `area-law' states are not difficult to engineer and do not look `thermal' at all. 
To illustrate this fact, let us consider a simple Hamiltonian with localized vertex terms and `hopping' interactions for each edge of the graph, 
\beq\label{hami}
H= \sum_{v\in\,{\rm vertices}} h_v + \sum_{\ell \in \,{\rm edges}} h_\ell\;,
\eeq
where $h_v$ acts on the single-vertex Hilbert space $\CH_v$ and the hopping term $h_\ell$ acts on the two-vertex Hilbert space $\CH_\ell=\CH_v \otimes \CH_w$ associated to the edge $\ell=(vw)$. Let us suppose now that $h_\ell \ll h_v$ and construct the ground state in the hopping expansion in powers of $\lambda$, a small parameter characterizing the matrix elements of $h_\ell$. To zeroth order, the vacuum is determined by the on-vertex vacua $|0\ket_v$ for each operator $h_v$, 
$$
|{\rm Vac} \ket = |\,\Pi \,\ket  + \CO(\lambda) \equiv \otimes_v \,|0\ket_v + \CO(\lambda)
\;,$$
a completely factorized state for which $S_A =0$ for any bi-partition. To leading order in the hopping terms the ground state has the form
$$
|{\rm Vac}\ket=  \alpha \,|\,\Pi\,\ket + \lambda \sum_{\ell\,\in \,{\rm edges}} |\,\ell\,\ket + \CO(\lambda^2)
$$
where the sum extends over all links $\ell= (vw)$ of the lattice $G$ and the state $|\ell \ket$ differs from the completely factorized on-site vacuum $|\Pi\ket$ by an excitation on the $\ell$-edge Hilbert space $\CH_\ell = \CH_v \otimes \CH_w$. It is orthogonal to the zeroth-order vacuum, $\bra\,\Pi\,|\,\ell\,\ket=0$, and its particular form is determined by  the matrix elements of $h_\ell$ through the usual formulae of first-order perturbation theory. Normalization sets the constant $\alpha$ to the value $\alpha = \sqrt{1-\lambda^2 N_\ell}$, with $N_\ell$ the total number of edges in $G$. 

Given now a bi-partition $G=A\cup {\bar A}$, we define the state
$$
|\lambda_{A^o} \ket = \sqrt{\alpha} \,|\,\Pi_{A^o}\,\ket + {\lambda \over \sqrt{\alpha}} \sum_{\ell\in A^o} |\,\ell\, \ket\;,
$$
where  the link sum includes only those edges lying entirely within the {\it interior} $A^o$ of the subgraph $A$ and $|\,\Pi_{A^o}\ket = \otimes_{v\in A^o} |0\ket_v$. Using this state we can write the vacuum to leading order in the hopping expansion as
$$
|{\rm Vac}\ket = |\,\lambda_{A^o} \,\ket \otimes |\,\lambda_{{\bar A}^o}\,\ket + \lambda\sum_{\ell_\pt} |\,\ell_\pt\,\ket + \CO(\lambda^2)\;,
$$
where $\ell_\pt$ denotes edges on the boundary $\pt A$.
In other words, the vacuum remains factorized except for  $\CO(\lambda)$ non-trivial excitations localized on the boundary of the partition. It follows that the entanglement entropy of the partition is proportional to the number of edges on the boundary, i.e.
$$
S_A =  \lambda\,c\,N_\ell (\pt A) + \CO(\lambda^2)\;,
$$
with $c$ a positive numerical constant.

Hence, we recover the expected result: ground states with very short-range correlations have `area-law' entanglement across bi-partitions. The peculiar behavior of expander graphs regarding Page-like criteria can be further emphasized by looking at the opposite limit, in which we regard the vertex hamiltonian density as negligible compared to the hopping term,  i.e. in the limit $|h_v | \ll |h_\ell |$. Let us assume that the Hamiltonian is homogeneous, in the sense that $h_\ell$ is given by the same operator for all links, so that $[h_\ell, h_{\ell'} ] =0$ for all links on $G$. Then, the evolution operator can be approximated as
\beq\label{simev}
U_G (t) = \prod_{\ell\,\in \,{\rm edges}} U_\ell\;, \qquad U_\ell = e^{-it h_\ell}\;,
\eeq
with $h_\ell$ independent of $\ell$, so that the ordering in the operator product is actually immaterial. If the initial state is still given by the
factorized ground state of the $h_v$ Hamiltonian, $\otimes_v |0\ket_v$, since in general $[h_v, h_\ell ] \neq 0$ when $v\in \ell$, we see that
each link operator $U_{vw}$ will generate entanglement in the $\CH_v \otimes \CH_w$ subspace in a time scale set by the eigenvalues of $h_\ell$. Since $h_\ell$ is interpreted as the hopping operator, we call $\beta$ the characteristic time scale for its inverse normal frequencies. 
The  resulting state with $\CO(1)$ entanglement across links  is generally termed a {\it graph state} in the  literature (cf. \cite{graphstates}), and its entanglement entropy is of the order of the rank of the off-diagonal adjacency sub-matrix connecting $A$ and ${\bar A}$. For a regular graph with fixed coordination, $k$, this so-called `Schmidt rank' is of the order of the
boundary edge-size, $N_\ell (\,\pt A\,)$, up to $k$-dependent degeneracy factors. 

We thus conclude that hopping-dominated hamiltonians on regular graphs achieve the {\it graph state} with area-law entanglement in times of $\CO(\beta)$. If the regular graph is also an expander, then we satisfy the Page criterion in `scrambling' times of $\CO(\beta)$. Such a ultra-fast scrambling time is undoubtedly a consequence of the highly integrable properties of the pure-hopping Hamiltonian with commuting link operators. For example, taking $\CH_v$ to be a two-level quantum system and $h_\ell$ determined by a single fundamental frequency of time scale $\beta$, the evolution operator (\ref{simev}) is actually periodic on time scales of $\CO(2\beta)$, so that it has very small times of Poincar\'e return, making it hard to talk physically about `scrambling' or any type of  thermalization for that matter. \footnote{The same can be said about the random Ising model example presented in \cite{haydennew}. In that case the scrambling time is of $\CO(\log \,N)$ because it takes $\CO(\log\,N)$ time steps to build a random lattice with $\CO(\log \,N)$ average vertex degree.}

It would be interesting to study  the scrambling times, in the sense of the Page criterion, for more generic Hamiltonians in the class (\ref{hami}), without the extreme simplifying assumptions of vertex and/or link domination. In any case, the comments offered in this section suggests that the Page criterion is not nearly enough to successfully characterize scrambling times in expander graphs.

\subsection{Going ballistic}

\noindent

Given the previous comments, it might be interesting to look for alternative characterizations of `fast' scrambling in expander graphs. One possibility is to focus attention on the `ballistic' character of expander-diffusion as the defining property, and look for quantum criteria adapted to this fact. The diffusion model based on random walks assumes the stability of the walking probe. This in turn is equivalent to the assumption that the system has a conserved quasiparticle density or, more generally,  a locally conserved current. On general grounds, retarded correlation functions of such local currents will show a characteristic `diffusion pole'  in the lower half plane of complexified frequencies. For local diffusion on an  Euclidean geometry with diffusion constant $D$, such a pole takes the form $\omega_{\rm diff} = -i\,D\,{\vec k}^{\,2} + \dots$ in a large wavelength expansion. Setting $|\,{\vec k}\,| \sim 1/L$ for a fluctuation of wavelength $L$, we find the standard diffusion time scale $D^{-1} L^2$ of  a slow scrambler.\footnote{This diffusion quasi normal mode is visible in the gravitational description of hydrodynamics, even if a quasiparticle picture is absent in this framework, provided $D$ is of $\CO(1)$ in the large $N$ limit.} 

According to the general discussion of section 2, the behavior of retarded correlation functions is expected to be quite different when the diffusion takes place in an expander graph. As a preliminary discussion, we may look at the continuum approximation and concentrate on the behavior of retarded correlation functions for field theories on hyperbolic manifolds. A first indication in this direction is the behavior of the free-field conformal Green's functions on hyperbolic space-times of the form ${\bf R} \times {\bf H}^{n}$, with the first factor representing a time-like dimension,\footnote{When comparing formulae with previous sections, notice that we have set $n=d+1$ in this subsection.} 
\beq\label{gf}
G_{\rm ret} (t,x) = \left\langle 0,0\,
{\Big |}\,\left( \pt_t^2  -\nabla^2_{{\bf H}^{n}} + {n-1 \over 4n} R_{{\bf H}^n}\right)^{-1}\; {\Big |} \,t, x\,\right\rangle_{\rm ret} \sim \delta(t-L_x)\; {1 \over \left(\sinh L_x\right)^{n-2}}\;, 
\eeq
where $(t,x) \in {\bf R} \times {\bf H}^{n}$  and $L_x$ is the hyperbolic geodesic distance from the conventionally chosen origin of coordinates on ${\bf H}^{n}$ to the point $x$ (cf. for example \cite{opticalnew}). We see that the retarded correlation function is indeed {\it ballistic}, since it is strictly supported over the past light-cone of the origin, despite the presence of a constant-curvature term which simulates a mass. This Green function is {\it automatically thermal}, since it is periodic under imaginary-time shifts $t\rightarrow t + 2\pi i$ when integrated against arbitrary sources. The effective temperature for a hyperboloid of radius $R$ is then $\beta^{-1} =(2\pi R)^{-1}$. 

The fact that perturbations on a free system behave ballistically at finite temperature is not surprising. After all, the retarded Green function of any free field theory is a c-number, being proportional to the commutator function of free fields, and thus it  is the same in any state. On the other hand, the functional form in (\ref{gf}) is largely dictated by conformal symmetry, and in particular the hyperbolic Green function must be
proportional to the {\it vacuum} Green function on Minkowski space, since ${\bf R}\times {\bf H}^{n}$ can be conformally mapped into a proper region
of ${\bf R}\times {\bf R}^{n}$, given by the causal development of the unit ball ${\bf B}^n$ on the $t=0$ spatial section ${\bf R}^n$. This region of Minkowski space,  denoted ${\cal D}({\bf B}^n)$,  has the form of a `diamond' based on ${\bf B}^n$.

This fact suggests a very interesting and non-trivial generalization of (\ref{gf}) to the case of fully interacting (but still conformal) field theories.  It is known (cf. for example \cite{casini}) that a CFT thermal state of temperature $1/2\pi$ on ${\bf R}\times {\bf H}^{n}$ is conformally related
to a vacuum state on ${\bf R}\times {\bf R}^{n}$ in the following sense:\footnote{Notice that a variant of  the same argument was used before to relate the Hartle--Hawking state on Rindler to the thermal state on the hyperboloid.} any time-dependent correlation function on the {\it hot} hyperboloid is conformally related to a {\it vacuum} Minkowski correlation function, {\it restricted} to ${\cal D}({\bf B}^n)$. 

Denoting $\eta_{\mu\nu}$ the Minkowski metric on ${\bf R}\times {\bf R}^{n}$ and $g_{\mu\nu}$ the hyperbolic metric on ${\bf R}\times {\bf H}^{n}$, we write
$\eta_{\mu\nu} = \Omega^2 \, g_{\mu\nu}$ for the conformal rescaling. It turns out that the whole hyperbolic space-time is mapped to the finite domain ${\cal D}({\bf B}^n)$ of Minkowski space-time. The Minkowski vacuum, as measured by local correlation functions restricted to ${\cal D}({\bf B}^n)$, is equivalent to the mixed
state obtained by tracing over the degrees of freedom `outside' the unit ball, i.e. we may replace the Minkowski vacuum by the entanglement density matrix on the unit ball. The conformal map sends this entanglement density matrix into the thermal density matrix
of the hyperbolic space-time. If the correlation functions are defined for scaling operators of conformal dimension $\Delta_i$ we have
\beq\label{confma}
\bra\,\CO_1 (y_1) \cdots \CO_s (y_s)\,\ket_{{\rm \beta},\,{\bf H}^n} = \Omega(y_1)^{\Delta_1} \cdots \Omega(y_s)^{\Delta_s} \,\bra\, {\bar \CO}_1 ({\bar y}_1) \cdots {\bar \CO}_s ({\bar y}_s)\,\ket_{{\rm vac}, \,{\cal D}({\bf B}^n)}
\eeq
where the  operators ${ \CO} ({ y})$ at points ${\bar y}\in {\bf R}\times {\bf H}^{n}$ are the images under the conformal map  of the operators  ${\bar \CO} ({\bar y})$  at points ${\bar y} \in {\cal D}({\bf B}^n)$. 

Consider now the retarded two-point function of an operator ${\bar \CO}({\bar y}) = {\bar \CO}_\Delta ({\bar t}, {\bar x})$ in the Minkowski vacuum of the CFT. We may write it as the imaginary part of the time-ordered correlation function, i.e. 
$$
{\bar G}_{\rm ret} ({\bar t}, {\bar x}) \propto \theta(-{\bar t}\,)\cdot {\rm Im}\,\left\langle \,T\left[ {\bar \CO}_\Delta (0, 0) \,{\bar \CO}_\Delta ({\bar t}, {\bar x})\right] \right\rangle_{{\rm vac},\, {\cal D}} = {\rm Im}\;{\theta(-{\bar t}\,) \over (-{\bar t}^{\,2} + |{\bar x}|^2 +i\,0)^{\Delta}}\;,
$$
where $\theta(t)$ is the Heaviside  function and we have used conformal invariance in the last step. Since we assume the CFT to be unitary, $\Delta$ is real, and the retarded correlation function is a distribution supported on the Minkowski  (past) light cone of the origin,  ${\bar t}^{\,2} = |{\bar x} |^2 $, ${\bar t} <0$,  just like it was for free fields.  Applying now the conformal map (\ref{confma}) we learn that the  {\it thermal} retarded Green function on the hyperbolic manifold
$$
G_{\rm ret} (t, x) = \Omega(t, x)^{\Delta} \;\Omega(0,0)^{\Delta} \;{\bar G}_{\rm ret}({\bar t}, {\bar x})
$$
is supported on the image of the light-cone under the conformal map. The conformal transformations send light-cones into light-cones, so that $G_{\rm ret} (t,x)$ is supported on the past light-cone of the origin in ${\bf R} \times {\bf H}^{n}$. This proves that the fully interacting Green function in the hyperbolic space-time is supported on the light-cone, and thus any diffusion poles (for $\CO_\Delta (y)$ equal to a conserved current of dimension $\Delta = n$) must be compatible with the ballistic propagation on the hyperboloid. 

It would be interesting to prove the ballistic character of thermal Green functions directly for discrete quantum systems on finite expander graphs. Perhaps such a proof can be approached by first solving the intermediate case of the Bethe lattice, a discrete expanding geometry of infinite volume.

\section{Concluding remarks}

\noindent

In this paper we have pointed out that a classic result in graph theory, namely the exponentially fast mixing of random walks on {\it expander graphs}, provides an interesting semi-phenomenological model for fast scrambling at black-hole horizons. 

Let us briefly recapitulate the logical steps leading to this conclusion. First, expander graphs are discrete versions of hyperbolic geometry, so that fast brownian mixing is akin to fast diffusion on hyperbolic spaces. On the other hand, regular horizons are conformal to asymptotically locally hyperbolic spaces,  the Hartle--Hawking quantum state being mapped to a homogeneous thermal ensemble on this so-called  `optical frame'. Working in this frame, it becomes then natural to identify the stretched horizon with a cut-off expander graph, hosting a quantum system with effective temperature $1/\beta$. This identification relates fast-scrambling to diffusion in the `optical depth' of the graph, turning into more standard `slow' scrambling when the bottom of the expander is reached. The relevant length scale separating the two scrambling modes is naturally given by the thermal length of the system.    

This geometrical picture of the stretched horizon also dovetails well with the chaotic-billiard model of `atmospheric' scattering developed in \cite{usuno}. While  the expander graph diffusion accounts for the scrambling of information absorbed by the stretched horizon, the billiard model accounts for the scrambling of information kept in the thermal atmosphere.

The most important open problem regarding these ideas is the {\it derivation} of the expander graph structure from a complete holographic description. In this regard, a natural expectation might be that the expander graph structure could survive the limit to weak coupling in the holographic dual, since the fast-scrambling scale, proportional to $\log N$, is not explicitly dependent on 't Hooft coupling factors. If this expectation is borne out, the fast-scrambling properties of sufficiently non-trivial matrix models (cf. \cite{matr}) could rely on identifying an expander graph structure in its Hilbert space, perhaps allowing for the expander graph to be defined as a random lattice.

 For this purpose, it is important to free the formal treatment   of scrambling from the simple random-walk model. A preliminary discussion along those lines is offered in the fourth section of this paper, where it is argued that standard formal criteria, such as the Page test, are not necessarily  sharp enough, and a more formal characterization in terms of ballistic behavior of thermal correlation functions is proposed. 
 
 It should be emphasized that the local operators referred to in section 4 are bulk objects, with an extremely non-local relationship to a given basis of local operators on the holographic dual. 
Indeed, the formulation of scrambling in the  `matrix-model', language  brings into focus the question of what observables are adequate. 
On general grounds, correlation functions of local operators in large-$N$ CFTs are expected to show quasi normal behavior with decay time of $\CO(\beta)$ and long-time quantum noise with infinite-time average of $\CO(\,e^{-S}\,)$ and associated Poincar\'e recurrences on time scales
of $\CO(\,\beta e^S\,)$ (cf. \cite{recs}). Quasinormal behavior is thus irrelevant for time scales longer than $\CO(\beta S)$, becoming then  subsumed into the quantum noise.  On the other hand, diffusion times of slow-scramblers can be detected in the correlation functions of locally conserved densities, associated to the standard diffusion poles in any hydrodynamical description. 

Detecting the $\log N$ time scale directly in CFT correlation functions is likely to require a non-trivial CFT definition of local operators, deep inside the near-horizon region.  How to characterize such extremely  bulk-local  operators remains a difficult challenge.   Other evidences for fast thermalization times are directly visible at $N=\infty$ in terms of non-local sharp probes, as in \cite{sharp}.

\vspace{0.5cm}
 
{\bf Acknowledgements:}  We are indebted to  E. Rabinovici for useful discussions. J.L.F.B. wishes to thank the KIPT at Santa Barbara for hospitality and the participants of the `Bits, branes and black holes' program for discussions. J.M.M. wishes to thank the theory group of Amsterdam University for hospitality and discussions. 
 The work of J.L.F.B. was partially supported by MEC and FEDER under a grant FPA2009-07908, the Spanish
Consolider-Ingenio 2010 Programme CPAN (CSD2007-00042) and  Comunidad Aut\'onoma de Madrid under grant HEPHACOS S2009/ESP-1473. J.M.M. is supported by a FPU fellowship from MICINN.

\end{document}